\documentclass[aps,prl,twocolumn,a4paper,superscriptaddress,nofootinbib]{revtex4-2}
\usepackage{graphicx,caption,subcaption,amsmath,amsfonts,amssymb,multirow,extarrows,bm,acronym,float,mathtools}
\usepackage{booktabs,multirow,tabularx}
\usepackage{xcolor}
\usepackage[colorlinks=true,citecolor=blue,linkcolor=blue,urlcolor=blue]{hyperref}
\usepackage[capitalize,nameinlink]{cleveref}
\graphicspath{{./Figures/}}
\newcommand{\Msun}{M_{\odot}}
\newcommand{\Hod}{\mathcal{H}}

\begin{document}
\title{Bekenstein--Hod Bound: A $3.3\sigma$ Confirmation from GW250114}

\author{Hai-Tian Wang}
\email{wanght9@dlut.edu.cn}
\affiliation{School of Physics, Dalian University of Technology, Dalian, Liaoning 116024, China}
\author{Shao-Peng Tang}
\affiliation{Key Laboratory of Dark Matter and Space Astronomy, Purple Mountain Observatory, Chinese Academy of Sciences, Nanjing, Jiangsu 210023, China}
\author{Yi-Zhong Fan}
\email{yzfan@pmo.ac.cn}
\affiliation{Key Laboratory of Dark Matter and Space Astronomy, Purple Mountain Observatory, Chinese Academy of Sciences, Nanjing, Jiangsu 210023, China}
\affiliation{School of Astronomy and Space Science, University of Science and Technology of China, Hefei, Anhui 230026, China}
\date{\today}

\begin{abstract}
Black holes link gravity, thermodynamics and information via limits on the relaxation rate of a perturbed system. The Bekenstein-Hod bound imposes a minimum relaxation time at fixed temperature, but its observational test demands both black hole thermodynamic characterization and decay time measurement.
Here we test this bound with GW250114, the 
loudest gravitational-wave signal yet observed from a binary black-hole merger. We infer the remnant temperature from pre-merger data truncated at least $10\,M$ before the peak and its longest-lived decay time from post-merger data, thereby avoiding direct reuse of the same strain samples. The ringdown frequencies and damping times are allowed to vary independently rather than being fixed to the Kerr spectrum. For the primary $t_{<}=-10\,M$ analysis, the bound is verified at $3.3-3.6\sigma$ across the focal ringdown start times, representing a substantial improvement over the $91\%$ confidence level set by GW150914. 
The conclusion remains robust under varied pre-merger cutoffs and explicit inclusion of the short-lived first overtone in waveform modelling. This separated-data measurement converts an information-theoretic relaxation bound into a precision test of a single astrophysical black hole.
\end{abstract}
\maketitle

\noindent\textit{Introduction---}\label{sec:intro}
Black holes provide an unusual setting in which gravity, thermodynamics and information theory make quantitative statements about the same physical system. The communication-theory foundations of information-rate limits trace to Shannon, with an early energy--time formulation due to Bremermann \cite{ShannonWeaver49,Bremermann}. Bekenstein then showed that causality and thermodynamics limit the rate at which a system with finite energy can transmit information \cite{Bekenstein1981}. Hod recast this statement as a time-times-temperature bound \cite{Hod2007}: a perturbed system at temperature $T$ cannot relax on a characteristic timescale shorter than $\hbar/(\pi k_{\rm B}T)$. Subsequent work examined the bound for Kerr--Newman and Reissner--Nordstr\"om black holes, its relation to geometric-optics estimates of quasinormal damping, and its statistical interpretation \cite{Hod:2008se,Hod:2009td,Hod:2018ifz,Ropotenko:2007jm}. Analyses of near-extremal Kerr black holes and their quasinormal spectra show how the limit can be approached in the appropriate regime \cite{Hod2007NearExtreme,Yang2013}, making black-hole relaxation a direct arena for testing the bound rather than a measurement of an information flux in bits per second.

A binary black-hole merger provides both observables needed for such a test. The inspiral and merger constrain the mass and spin of the remnant and hence, under the Kerr--Hawking relation, its temperature  \cite{Hawking1975}. The post-merger ringdown is described by damped quasinormal modes whose complex frequencies encode the remnant properties, as established in foundational perturbation studies and subsequent reviews  \cite{Vishveshwara,Press:1971wr,Chandrasekhar:1975zza,Kokkotas1999,Berti2009}. The relevant relaxation timescale is that of the longest-lived component governing late-time decay, not the shorter lifetime of every overtone. The detection and parameter inference of GW150914 opened this regime observationally  \cite{GW150914,PE_paper_GW150914,TGR_paper_GW150914}. Time-domain studies then established the accessibility of ringdown spectroscopy and the importance---and modelling sensitivity---of overtones near the signal peak  \cite{Carullo:2018sfu,PhysRevD.99.123029,Isi2019,Giesler2019,Bhagwat:2019dtm,Forteza:2020hbw,Okounkova:2020vwu,Wang2023}. Carullo and colleagues first combined remnant and ringdown measurements for a Bekenstein--Hod test and found 91\% support from GW150914 and 94\% population-level support for the bound using events identified in the first two gravitational-wave transient catalogues and their strong-field tests  \cite{GWTC-1,GWTC-2,GWTC-2_TGR,Carullo2021}. 

GW250114 is exceptionally suited to a separated-data test. Its network signal-to-noise ratio is approximately 76--80, depending on the analysis convention, and the signal supports precise inference both before and after merger  \cite{Abac2025,Abac2026}. The LIGO--Virgo--KAGRA (LVK) analyses found that the post-merger data require at least two damped components; the more rapidly damped component is consistent with the Kerr $(2,2,1)$ overtone for about one oscillation cycle  \cite{Abac2026}. 
The exceptionally clean post-merger signal has further been utilized to search for signatures of the perturbed remnant horizon  \cite{Lu:2025vol}.
The same event also enabled a stringent test of Hawking's area law using independently truncated pre- and post-merger data \cite{Hawking1971,Abac2025,Abac2026}, consistent with independent tests using GW230814 \cite{Tang:2025jyj} and population-level analyses \cite{Tang:2026ajp}. The area law constrains the global change of horizon area, whereas the Bekenstein--Hod inequality constrains the dynamical rate of relaxation, so the two tests probe distinct aspects of black-hole thermodynamics.

For a Kerr black hole of detector-frame mass $M_f$ and dimensionless spin $\chi_f$, the Hawking temperature is \cite{Hawking1975}
\begin{equation}
 T_{\rm H}=\frac{\hbar c^3}{4\pi G k_{\rm B}M_f}
 \frac{\sqrt{1-\chi_f^2}}{1+\sqrt{1-\chi_f^2}}.
 \label{eq:hawking-temperature}
\end{equation}
Our central observable is
\begin{equation}
 \Hod \equiv \frac{\tau_{\rm BH}}{\tau_{220}}=\frac{\hbar}{\pi k_{\rm B}T_{\rm H}\tau_{220}},
 \qquad \Hod\leq 1,
 \label{eq:hod-definition}
\end{equation}
where $\tau_{220}$ is the damping time of the longest-lived recovered component.

Eqs.~(\ref{eq:hod-definition}) and (\ref{eq:hawking-temperature}) give
\begin{equation}
 \Hod=\frac{4GM_f}{c^3\tau_{220}}
 \frac{1+\sqrt{1-\chi_f^2}}{\sqrt{1-\chi_f^2}}.
 \label{eq:hod-observable}
\end{equation}
Both $M_f$ and $\tau_{220}$ are evaluated in the detector frame, so the common cosmological redshift cancels. We obtain $(M_f,\chi_f)$ from three inspiral--merger (IM) analyses whose data are truncated before the peak, and obtain $\tau_{220}$ from ringdown analyses whose data begin at a chosen time after the peak. Posterior samples from the two sides are paired independently when constructing $\Hod$.

\begin{figure*}[t]
\centering
\includegraphics[width=0.96\textwidth]{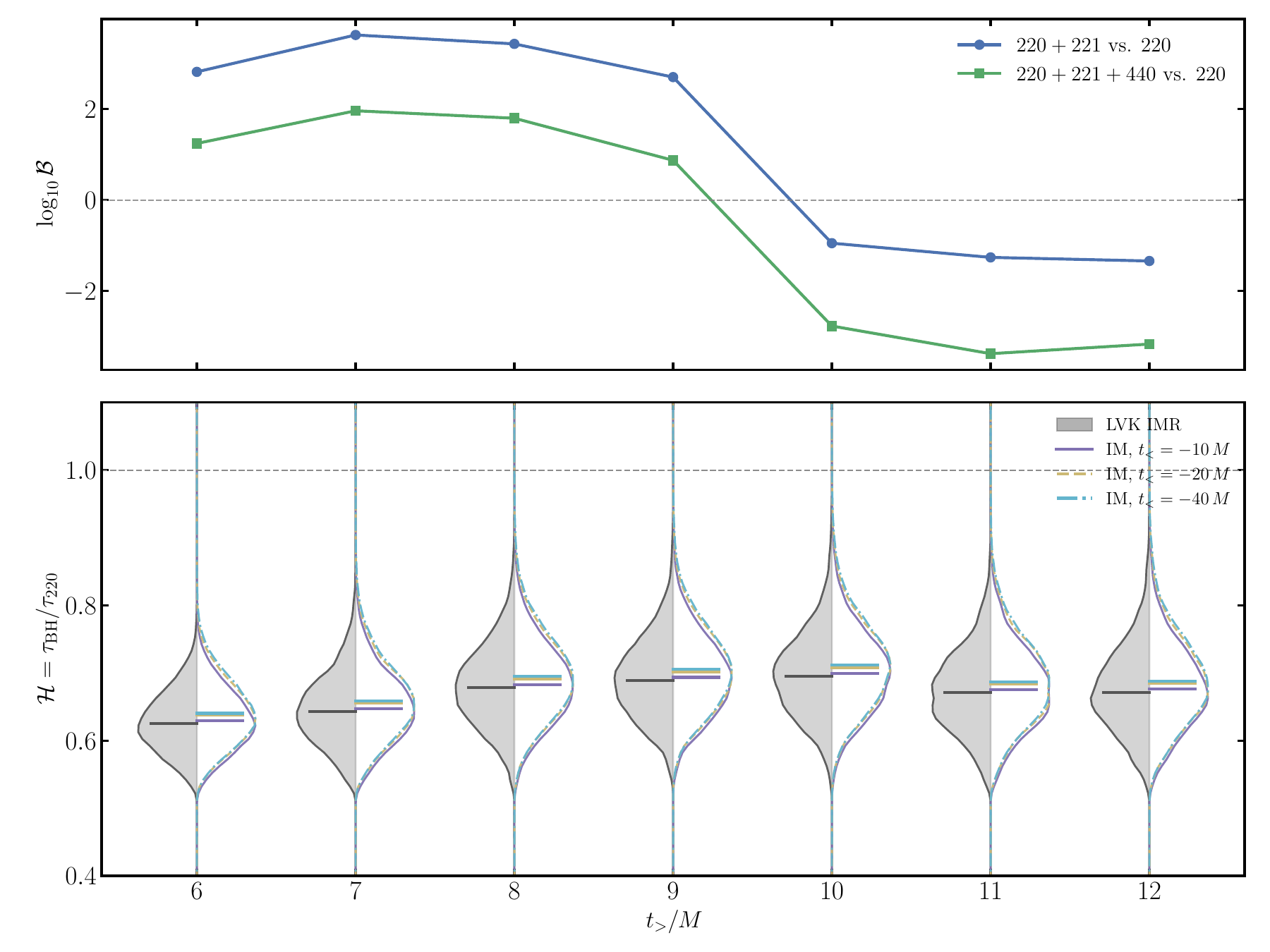}
\caption{Top, base-10 logarithm of the Bayes factor for the freely varying $220+221$ model (blue) and $220+221+440$ model (green), each relative to the single-$220$ model, as a function of ringdown start time after the polarization peak. Bottom, posterior distributions of $\Hod$ obtained by combining the single-$220$ damping-time posterior with the LVK full-IMR remnant posterior (left half of each violin) or with remnant posteriors from IM data truncated at $-10\,M$, $-20\,M$ and $-40\,M$ (right half). Horizontal bars mark posterior medians. The dashed line denotes the bound $\Hod=1$.}
\label{fig:bound}
\end{figure*}

\noindent\textit{Setting---}
We analysed publicly released Hanford and Livingston strain for GW250114 and the LVK analysis products associated with the discovery paper \cite{GWOSC2025,LVKData2025}. The reference polarization-peak time was GPS $t_{\rm peak}=1420878141.235932$. Time offsets were expressed using $M=68.1\,\Msun$, for which $1M=GM/c^3\simeq0.34$\,ms in the detector frame. The reference sky position used to define the nominal detector arrival times was right ascension $\alpha_0=2.33$\,rad and declination $\delta_0=0.19$\,rad; the polarization angle was $\psi_0=1.33$\,rad. In the primary analyses, right ascension, declination and geocentric time were sampled rather than fixed.

The raw 4096-Hz strain was downsampled to 2048\,Hz using a Butterworth anti-alias filter and high-pass filtered at 20\,Hz with a finite-impulse-response filter. We used the one-sided power spectral densities released with the LVK posterior products. The noise autocovariance was obtained from the inverse Fourier transform of each power spectral density, following the time-domain procedure validated in previous analyses \cite{Wang2023,Wang2024}. Ringdown segments were 0.4\,s long; pre-merger segments were 1.4\,s long, matching the LVK GW250114 setup \cite{Abac2025}. For every analysis, the truncation applied to the detector data vector and hence to the noise log likelihood was evaluated with the detector-specific peak times fixed according to the pre-fixed sky location ($\alpha_0$, $\delta_0$) and reference polarization-peak time $t_{\rm peak}$ described above. Only the waveform truncation entering likelihood sampling varied with the sampled geocentric time and sky-dependent detector delays.

We define time offsets relative to the polarization peak time $t_{\rm peak}$ as $t_{<,>} = t - t_{\rm peak}$, with $t_{<} < 0$ denoting the pre-merger truncation time prior to the peak and $t_{>} > 0$ denoting the ringdown analysis start time following it. We therefore infer the remnant temperature from strain ending at $t_{<} = -10\,M$, $-20\,M$ or $-40\,M$, where $M=68.1\,\Msun$ is the detector-frame remnant-mass scale used to convert time offsets, and infer the relaxation time from ringdown strain starting at $t_{>} \geq 6\,M$.
The three pre-merger cutoffs form a sensitivity scan over the amount of excluded merger data; the $-40\,M$ cutoff is included as the most conservative separation. For the ringdown, we fit freely varying frequencies and damping times, imposing no Kerr spectral relation. 

\begin{figure*}[t]
\centering
\includegraphics[width=0.80\textwidth]{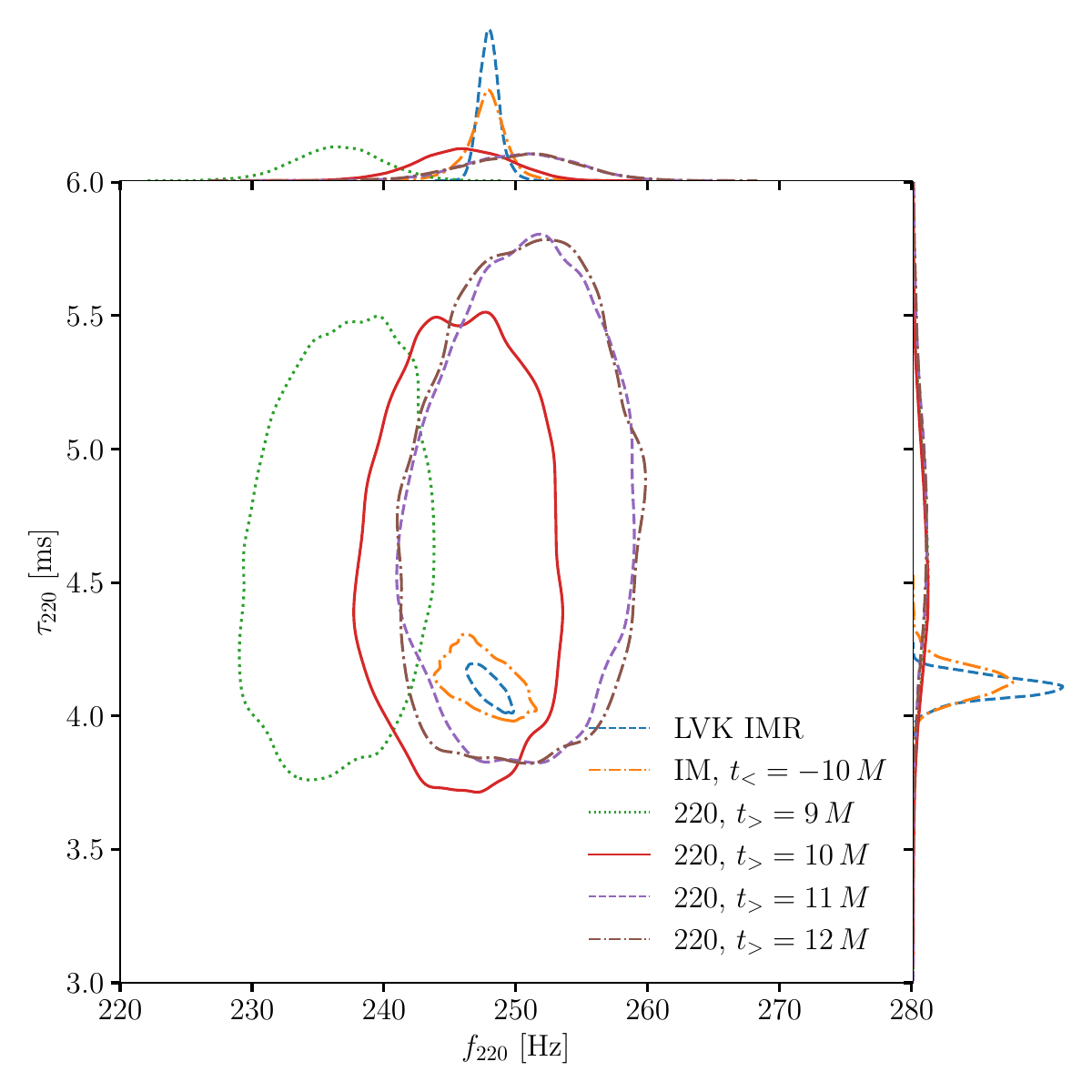}
\caption{Joint 90\% credible contours for the detector-frame $220$ frequency and damping time. Green, red, purple and brown contours show single-mode ringdown analyses beginning at $9\,M$, $10\,M$, $11\,M$ and $12\,M$, respectively. Blue and orange contours are Kerr predictions obtained from the released LVK full-IMR posterior and the independently truncated IM posterior with $t_{<}=-10\,M$. The Kerr relation is used only to draw the comparison contours, not in the ringdown likelihood.}
\label{fig:spectrum}
\end{figure*}

The resulting probability densities are shown in FIG.~\ref{fig:bound}. Using the released LVK full inspiral--merger--ringdown (IMR) remnant posterior gives a narrower reference result but reuses post-merger information and is not our primary inference. The three truncated IM results are mutually consistent and slightly broader, as expected when progressively more high-amplitude data are discarded. This modest loss of precision provides a clean separation between the strain samples determining the temperature and those determining the relaxation time.

\noindent\textit{Ringdown model selection---}
At early start times, the post-merger data favour modelling the short-lived component. Relative to a single $220$ damped sinusoid, the $220+221$ model has $\log_{10}\mathcal{B}=2.82$, $3.64$, $3.44$ and $2.71$ at start times $6\,M$, $7\,M$, $8\,M$ and $9\,M$, respectively. The Bayes factor becomes negative at $10\,M$ and remains so at later times, showing that the overtone contribution can be neglected for the late-time relaxation measurement. This transition is consistent with the LVK finding that the second, more rapidly damped component remains informative for only about one cycle \cite{Abac2026}.

\begin{table*}[t]
\centering
\caption{Fundamental-mode constraints and Bekenstein--Hod significance. Values are posterior medians with central 90\% credible intervals. The last four columns give $X_{\rm BHU}$ after combining each ringdown damping-time posterior with the indicated remnant posterior. The significance is computed from the separation of the means relative to the quadrature sum of the posterior standard deviations.
}
\label{tab:spectrum}
\begingroup
\small
\setlength{\tabcolsep}{3.5pt}
\begin{tabular}{lcccccc}
\toprule
\multirow{2}{*}{Ringdown} & \multirow{2}{*}{$f_{220}$ [Hz]} & \multirow{2}{*}{$\tau_{220}$ [ms]} & \multicolumn{4}{c}{$X_{\rm BHU}$ [$\sigma$]} \\
\cmidrule(lr){4-7}
 & & & LVK IMR & \shortstack{IM\\$t_{<}=-10\,M$} & \shortstack{IM\\$t_{<}=-20\,M$} & \shortstack{IM\\$t_{<}=-40\,M$} \\
\midrule
$t_{>}=9\,M$ & $236.51_{-5.46}^{+5.39}$ & $4.58_{-0.61}^{+0.69}$ & $3.60$ & $3.52$ & $3.41$ & $3.32$ \\
$t_{>}=10\,M$ & $246.03_{-6.03}^{+5.54}$ & $4.55_{-0.62}^{+0.71}$ & $3.42$ & $3.34$ & $3.23$ & $3.15$ \\
$t_{>}=11\,M$ & $250.53_{-6.87}^{+6.40}$ & $4.71_{-0.67}^{+0.78}$ & $3.63$ & $3.56$ & $3.45$ & $3.37$ \\
$t_{>}=12\,M$ & $250.56_{-7.38}^{+6.60}$ & $4.71_{-0.69}^{+0.79}$ & $3.52$ & $3.45$ & $3.35$ & $3.28$ \\
\bottomrule
\end{tabular}
\endgroup
\end{table*}

Adding a freely varying $440$ component is not supported. The $220+221+440$ model is disfavoured relative to the $220+221$ model at all tested start times (FIG.~\ref{fig:bound}). This comparison uses an agnostic spectroscopy model that assigns an independent frequency, damping time, amplitude and phase to every component, rather than mapping the mode spectrum to a shared Kerr remnant mass and spin. Because the data do not require the $440$ component, it is excluded from the Bekenstein--Hod inference.

These comparisons define two complementary regimes. At $6\,M$--$8\,M$, the $220+221$ model is used to check that the fundamental can be recovered without absorbing the overtone. At $9\,M$--$12\,M$, we focus on the freely varying single-$220$ model, with $10\,M$ and later providing the cleanest model-selection support for a fundamental-only description. The universal relaxation bound is applied only to $\tau_{220}$; the faster $221$ decay is a nuisance component at early times, not an independent candidate for violating the bound.

\noindent\textit{The bound across analysis choices---}
All combinations of the four primary ringdown start times and three pre-merger truncations support $\Hod<1$. Following the significance convention used in the LVK area-law analysis of GW250114 \cite{Abac2026}, we quantify the separation of the independently inferred relaxation time $\tau_{220}$ and Bekenstein--Hod timescale $\tau_{\rm BH}$ using their means and variances. For the primary truncation, $t_{<}=-10\,M$, the posterior medians range from 0.676 to 0.700 and the bound is satisfied at $3.34$--$3.56\sigma$ over $9\,M$--$12\,M$. Moving the pre-merger cutoff to $-20\,M$ or $-40\,M$ broadens the remnant prediction but changes neither its location qualitatively nor the conclusion. The smallest significance among these robustness choices is $3.15\sigma$, obtained for a $10\,M$ ringdown start and a $-40\,M$ pre-merger cutoff; this configuration gives $\Hod=0.712$ with a central 90\% credible interval of $[0.610,0.832]$ and $P(\Hod<1)=99.972\%$. Repeating both the pre-merger and ringdown inference with the sky location and geocentric time fixed to their reference values changes the numerical posteriors modestly but leaves the conclusion unchanged.


For the released LVK full-IMR posterior, the bound is satisfied at $3.42$--$3.63\sigma$ over the same start times. The truncated IM analyses are more important physically: they show that the strong support does not depend on remnant information extracted from the post-merger strain.

\noindent\textit{Spectral consistency---}
The freely inferred late-time spectra can be compared with Kerr frequencies and damping times predicted from the remnant posteriors without imposing that prediction in the ringdown likelihood (FIG.~\ref{fig:spectrum} and Table~\ref{tab:spectrum}). At $9\,M$, the recovered frequency lies below the narrow Kerr prediction, whereas its damping-time posterior remains compatible. From $10\,M$ onward, the joint posterior moves towards the predicted fundamental spectrum. The broad ringdown uncertainties dominate the uncertainty in $\Hod$; truncating the pre-merger data mainly broadens the already precise remnant prediction.

\noindent\textit{Discussion---}
GW250114 provides a qualitatively cleaner Bekenstein--Hod test than was possible with earlier gravitational-wave events. Its unusually informative pre- and post-merger signals allow the two factors in Eq.~(\ref{eq:hod-definition}) to be inferred from non-overlapping data. The agreement across $-10\,M$, $-20\,M$ and $-40\,M$ truncations shows that the conclusion is not driven by strain samples near the peak, while the ringdown model comparison isolates the epoch at which the longest-lived component can be measured without a required overtone.

The result is a consistency test with a specific theoretical interpretation. The damping time is measured without enforcing the Kerr QNM spectrum, but conversion of the pre-merger remnant mass and spin into a temperature assumes the Kerr--Hawking relation. The measurement is therefore not a direct detection of Hawking radiation and does not measure an emitted information rate in bits per second. Rather, it tests the dimensionless time--temperature combination through which the information-rate argument constrains macroscopic relaxation.

Every data-separated configuration places the posterior well below $\Hod=1$ and satisfies the bound at more than $3\sigma$, with the conclusion unchanged across the tested pre-merger cutoffs, ringdown start times and extrinsic-parameter treatments. The reported $X_{\rm BHU}$ values summarize the separation through the first two posterior moments.
The present result is a precision single-event test, with its scope set by the adopted waveform and noise models.

More broadly, this result brings an information-theoretic limit on relaxation into the same observational setting as tests of strong-field gravity and black-hole thermodynamics. Its significance is that the separated-data construction provides an empirical benchmark for the allowed relation between a black hole's thermodynamic temperature and its macroscopic decay time. 

The calibrated GW250114 strain is available from the Gravitational Wave Open Science Center \cite{GWOSC2025}. LVK posterior samples are available from Zenodo \cite{LVKData2025}. 

This work made use of the following software packages: \texttt{astropy} \cite{astropy:2013, astropy:2018, astropy:2022}, \texttt{matplotlib} \cite{Hunter:2007}, \texttt{numpy} \cite{numpy}, \texttt{python} \cite{python}, \texttt{scipy} \cite{2020SciPy-NMeth, scipy_12522488}, \texttt{Bilby} \cite{Ashton:2018jfp, Bilby_2602178}, \texttt{PyCBC} \cite{Allen:2005fk}, \texttt{corner} \cite{corner-Foreman-Mackey-2016, corner.py_14209694}, \texttt{h5py} \cite{collette_python_hdf5_2014, h5py_7560547}, \texttt{pandas} \cite{mckinney-proc-scipy-2010, pandas_13819579}, \texttt{Jupyter} \cite{2007CSE.....9c..21P, kluyver2016jupyter}, \texttt{qnm} \cite{Stein:2019mop}, \texttt{lalsuite} \cite{lalsuite}, \texttt{swiglal} \cite{swiglal}, \texttt{GWpy} \cite{gwpy,Macleod:2021goi}, \texttt{pykerr} \cite{collin_capano_2023_10056495,Leaver:1985ax,Berti:2005ys}, and \texttt{sxs} \cite{Boyle_The_sxs_package_2025}.

\noindent\textit{Acknowledgements---}
This work was supported by the Natural Science Foundation of China under grant numbers 12603066, 12233011, and 12303056. H.-T. Wang is also supported by the Natural Science Foundation of Liaoning Province (Doctoral Scientific Research Start-up Fund, Grant No. 2025-BS-0065) and the Fundamental Research Funds for the Central Universities at Dalian University of Technology.
This research made use of data and software obtained from the Gravitational Wave Open Science Center, a service of LIGO Laboratory, the LIGO Scientific Collaboration, the Virgo Collaboration and KAGRA.

\bibliographystyle{apsrev4-2}
\bibliography{submit_arxiv,softs}
\clearpage
\onecolumngrid
\appendix
\setcounter{secnumdepth}{2}

\section{Methods}\label{sec:method}
\subsection{Pre-merger inference}
We performed three time-domain IM parameter-estimation analyses with the NRSur7dq4 waveform model \cite{Varma2019}. The data likelihood was truncated at $t_{<}=-10\,M$, $-20\,M$ or $-40\,M$ relative to the reference polarization peak, while all other data settings were held fixed. The waveform sample rate was 2048\,Hz and the low-frequency cutoff was 20\,Hz. The geocentric time prior was Gaussian with a standard deviation of 0.01\,s and was centered on $t_{\rm peak}$. Right ascension and declination were sampled subject to the requirement that, after accounting for the response delay of each detector, the sampled waveform peak lay within $-3\,M$ to $+3\,M$ of the corresponding reference peak. This constraint prevents sampling from becoming trapped in the secondary sky--time mode reported for the time-domain inspiral analysis of GW250114, for which the LVK analysis instead extended the Markov-chain burn-in \cite{Abac2025}. We also repeated the analyses with geocentric time and sky position fixed to their reference values to confirm that our results are robust.

For every posterior sample, component masses and spin coordinates were converted to Cartesian spins at a reference frequency of 20\,Hz. The NRSur7dq4 remnant fits were then used to obtain the detector-frame final mass $M_f$ and dimensionless spin $\chi_f$. The released LVK full-IMR NRSur7dq4 remnant posterior was processed in the same detector-frame convention and used only as a comparison.

\subsection{Ringdown model}
For $t\geq t_{>}$, the complex strain was modeled as a sum of damped sinusoids,
\begin{align}
 h_{+}(t)+ih_{\times}(t)
 &=\sum_{\ell mn}A_{\ell mn}e^{-(t-t_{>})/\tau_{\ell mn}} \notag\\
 &\quad\times e^{i[2\pi f_{\ell mn}(t-t_{>})+\phi_{\ell mn}]}
 {}_{-2}Y_{\ell m}(\iota,\delta).
\end{align}
We approximated spin-weighted spheroidal harmonics by spin-weighted spherical harmonics; the resulting mode mixing has been quantified for perturbed Kerr black holes \cite{Berti2014}. Crucially, $f_{\ell mn}$ and $\tau_{\ell mn}$ were sampled independently for every component; no remnant mass, spin or Kerr spectral relation entered the ringdown likelihood. The tested models contained $220$, $220+221$ or $220+221+440$ components. Ringdown start times ranged from $6\,M$ to $12\,M$ after the reference peak in steps of $1\,M$. As in the pre-merger analyses, the detector data truncation was fixed, whereas waveform truncation followed the sampled geocentric time and detector delays. We used the same 0.01-s Gaussian geocentric-time prior and the same per-detector $\pm3\,M$ peak-time constraint on the sampled sky location. Fixed-extrinsic-parameter runs provided the corresponding robustness check.

Frequencies had uniform priors of 150--300\,Hz for $f_{220}$, 200--300\,Hz for $f_{221}$ and 450--650\,Hz for $f_{440}$. Damping-time priors were uniform over 2.5--6\,ms for the $220$ and $440$ components and 0.5--2.5\,ms for the $221$ component. Each phase was uniform over $[0,2\pi]$, and each amplitude was uniform over $[0,5\times10^{-19}]$.

\subsection{Time-domain likelihood and sampling}
For detector data vector $\mathbf d$, model $\mathbf h(\Theta)$ and noise covariance matrix $\mathbf C$, the log likelihood was
\begin{equation}
 \ln\mathcal L(\Theta)=-\frac{1}{2}
 [\mathbf d-\mathbf h(\Theta)]^{\intercal}\mathbf C^{-1}
 [\mathbf d-\mathbf h(\Theta)]+C_0,
\end{equation}
where $C_0$ is a normalization constant and $\intercal$ denotes the transpose. The inner product $\langle \cdot | \cdot \rangle$ incorporates the noise characteristics.
All analyses used {\sc Bilby} package~\cite[v2.8.0;][]{Ashton2019} and {\sc Dynesty} \cite[v3.0.0;][]{Speagle2020}, with 1,000 live points, the acceptance-walk proposal, \texttt{naccept=60}. Bayes factors were calculated from the nested-sampling evidences using the single-$220$ model as the reference.

\subsection{Bekenstein--Hod posterior}
For each ringdown start time, we drew $10^6$ independent pairs from the $\tau_{220}$ posterior and from each $(M_f,\chi_f)$ remnant posterior and evaluated Eq.~(\ref{eq:hod-observable}); these draws determine the reported medians, central 90\% credible intervals, plotted densities and $P(\Hod<1)$. Because the remnant and ringdown likelihoods use non-overlapping data, their posterior samples were paired independently while preserving the mass--spin correlation within every remnant sample.

Following the convention adopted for the GW250114 area-law analysis \cite{Abac2026}, we report the significance of satisfying the Bekenstein--Hod bound as
\begin{equation}
 X_{\rm BHU}=\frac{\mu_{\tau_{220}}-\mu_{\tau_{\rm BH}}}
 {\sqrt{\sigma_{\tau_{220}}^2+\sigma_{\tau_{\rm BH}}^2}},
 \label{eq:bhu-significance}
\end{equation}
where $\mu_{\tau_{220}}$ and $\sigma_{\tau_{220}}$ are the mean and standard deviation of the freely inferred damping-time posterior, and $\mu_{\tau_{\rm BH}}$ and $\sigma_{\tau_{\rm BH}}$ are those of the Bekenstein--Hod timescale posterior calculated from $(M_f,\chi_f)$. The two variances are added because the pre- and post-merger measurements are independent. This moment-based statistic avoids estimating significance from sparsely sampled posterior tails beyond approximately $4\sigma$.

\section{Early-time two-component spectra}

\begin{figure*}[t]
\centering
\includegraphics[width=0.80\textwidth]{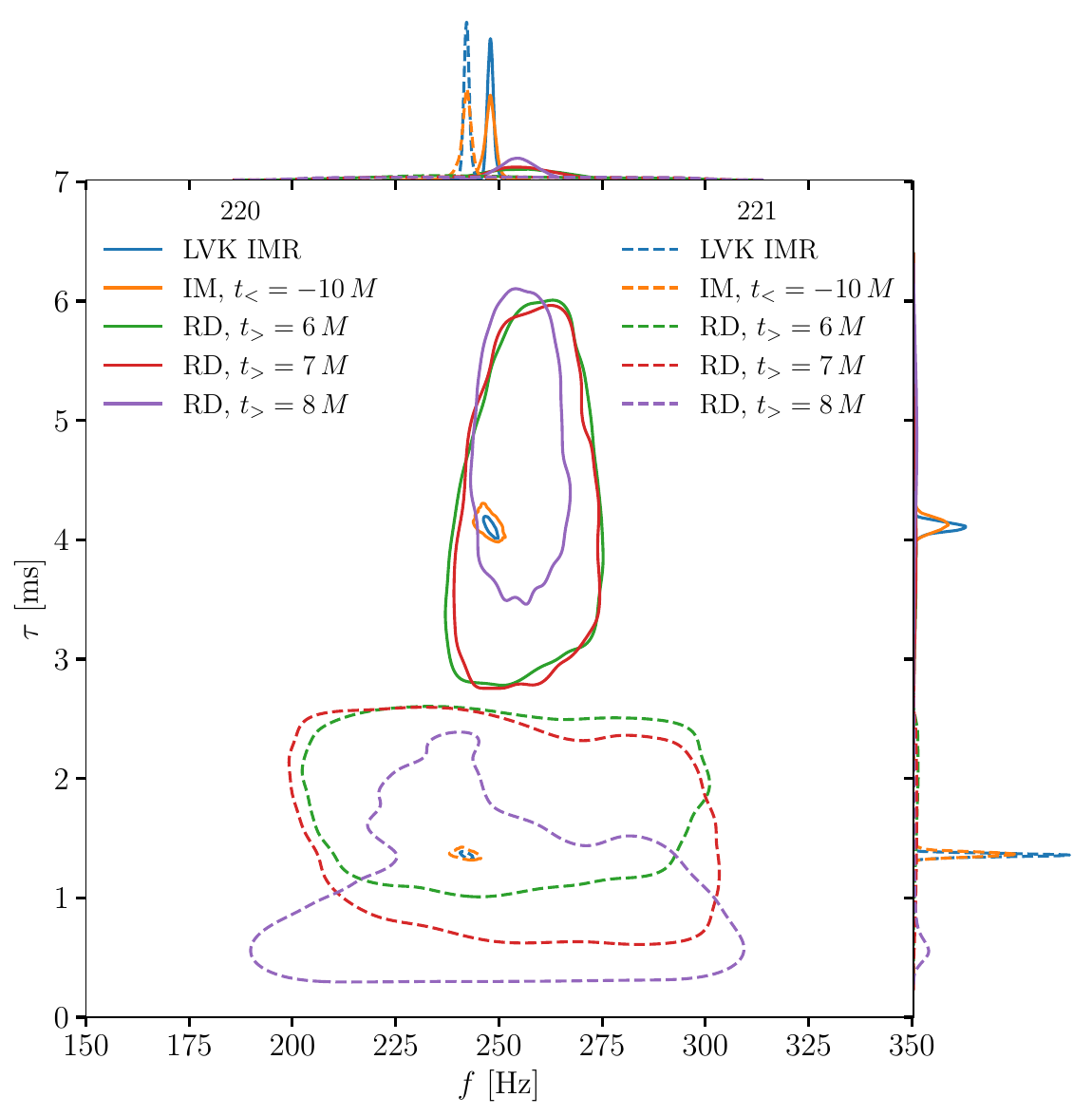}
\caption{Joint 90\% credible contours for freely varying $220$ (solid) and $221$ (dashed) frequencies and damping times at ringdown start times $6\,M$, $7\,M$ and $8\,M$. Blue and orange contours show Kerr predictions from the LVK full-IMR posterior and the independently truncated IM posterior with $t_{<}=-10\,M$. 
}
\label{fig:extended-two-mode}
\end{figure*}

FIG.~\ref{fig:extended-two-mode} presents a two-mode ringdown-spectroscopy analysis in which the frequencies and damping times of the $220$ and $221$ components are inferred freely, without imposing the Kerr hypothesis. By contrast, the LVK IMR and IM results (the latter obtained with different choices of $t_{<}$) are converted into the corresponding $220$ and $221$ spectra from their inferred final mass and final spin under the Kerr hypothesis. As the ringdown start time $t_{>}$ is increased, the inferred $221$ spectrum shifts progressively towards zero, most clearly in the posterior for $\tau_{221}$, consistent with the expectation that this rapidly damped component contributes less at later times.

\end{document}